\documentclass[12pt]{article}
\begin{document}

\title{$D$ and $J/\psi$ production from deconfined matter  \\
       in relativistic heavy ion collisions }
\author{ P. Csizmadia$^{1,2}$,  P. L\'evai$^{1,2}$\\[1ex]
$^{1}$ RMKI Research Institute for Particle and Nuclear Physics, \\
\ \  P. O. Box 49, Budapest, 1525, Hungary \\
$^{2}$ Physics Dept., Columbia University, New York, NY, 10027, USA
}

\date{August 18.   2000}

\maketitle
\begin{abstract}
We study the production of the $D$ and $J/\psi$ mesons from  
deconfined quark matter at CERN SPS energy.
Using the MICOR microscopical coalescence model we determine
the transverse momentum spectra of these charm mesons.
We predict the slopes of their transverse momentum spectra
in Pb+Pb collision at 158 GeV/nucleon beam energy.
\end{abstract}

During the last decade charm hadron production
was investigated with great interest
at the CERN Super Proton Synchrotron (SPS), because
the theoretically predicted anomalous suppression of the
$J/\psi$ particle  could
indicate the appearance of colour deconfinement and the formation
of the quark-gluon plasma (QGP) phase \cite{MatSat86}.
An anomalous suppression has been detected in Pb+Pb collisions
at 158 GeV/nucleon bombarding energy \cite{NA50supr} and this
result is considered as the strongest evidence of QGP formation 
(for an overview see Ref.~\cite{Satz00}).
However, the unambiguous confirmation of the QGP formation requires more 
charm related data, e.g.
D-meson spectra and absolute yields. This provides motivation 
to measure open charm production in Pb+Pb collision
at the CERN SPS \cite{letterNA60}.   
In parallel, new theoretical efforts work to 
determine charm hadron yields and ratios in this energy range
using statistical \cite{Gazd99},
thermal \cite{Munz00} and non-equilibrium \cite{ALCORC} models of
charm hadron production. 

Since momentum spectra could reveal some interesting 
details about the formation mechanisms of
charm hadrons and thus the properties of the early (supposedly deconfined) 
phase,  our aim is to determine the transverse momentum spectra 
of the $D$ and $J/\psi$ mesons in heavy ion collisions. We assume
that hadrons are produced directly from a deconfined state
by quark coalescence.
We neglect hadronic rescatterings. 
This hadronization scenario is modelled using
MIcroscopical COalescence Rehadronisation (MICOR)
\cite{MICOR00,MICORSQM98,MICORUDQ99}
which has been successfully applied to reproduce the measured transverse 
slopes of the $\phi$, $\Omega$ and $\rho$ mesons \cite{MICOR00}.
(The spectra of other light and strange hadrons were investigated 
previously in Refs.~\cite{MICORSQM98,MICORUDQ99}.)
The MICOR model is the successor of the ALCOR \cite{ALCOR1,ALCOR2}
and the Transchemistry \cite{Transchem} models. While the
ALCOR and Transchemistry
models were constructed to determine the total number of
hadrons produced from quark matter, 
the MICOR model can also determine the full momentum spectra of the
produced hadrons.  We therefore include charm quarks into the MICOR model
and calculate the spectra of directly produced charm mesons.

In the MICOR model we start with a  deconfined initial
phase as motivated by lattice QCD calculations. (Note that phase differs 
from the idealized, asymptotically free quark-gluon plasma \cite{Boyd95}.) 
The lattice results for the QCD equation of state (with $N_f=0,2,4$ flavour)
were parametrized phenomenologically as a massive quark matter 
\cite{LevHein}.  This parametrization 
reproduces the lattice QCD results regarding the pressure and energy density
in the temperature region $1 \leq T/T_c \leq 3$ \cite{LevHein} and thus it
can be used as a suitable description of hadronization.
The massive gluons
(whose number is suppressed relative to the quarks and antiquarks)
create an attractive effective
force between the colored quarks and anti-quarks which drives
the hadronization.  According to our assumption, the quark matter here is  
a fully thermalized state, and in the hadronization stage
is characterized by a hadronization temperature $T_{had}$,
a transverse flow $v_T^0$ and a Bjorken-scaled longitudinal flow.

The coalescence of the massive quarks produces a
hadron resonance gas, which is out of equilibrium
\cite{MICOR00}. This out of equilibrium character will not change during
the resonance decay.
While the final state hadronic spectra can be fitted by 
exponential functions on transverse momenta, this does not mean that
the obtained slope parameters characterize equilibrated hadronic 
gas states. 

During hadronization meson-like objects are formed  
via quark-antiquark coalescence
(see Ref.~\cite{ALCOR1} for the coalescence cross section), 
while baryon-like objects are formed 
in two steps: the formation of diquarks from quarks are followed by
the coalescence of diquarks and quarks into baryons.
We assume that color-neutral prehadrons escape the deconfined state
without desintegration and that they 
become part of the produced excited hadronic gas.
These prehadrons are off-shell, however their off-shell
masses are very close to the well-known bare hadron masses.
This minor difference is corrected in a last step of the hadronization,
when a prehadron  
absorbs or radiates the appropriate (small) energy to become on-shell.
We assume that it conserves its velocity during this process.
More details about the physics of the MICOR model can be found
in Ref.~\cite{MICOR00}.

The initial condition of the MICOR hadronization model consists of
the hadronization temperature, $T_{had}$, 
and the initial transverse flow, $v_T^0$, of the deconfined phase.
The analysis of the measured transverse momentum spectra of 
$\phi$, $\Omega$ and $\rho$ particles was performed in Ref.~\cite{MICOR00}
and fits to these  experiment data favor  initial values of
$T_{had} = 175 \pm 15$ MeV and of 
$v_T^0 = 0.46 \pm 0.05$ in the Pb+Pb collision.
We will use the mean values of these parameters to calculate 
charm meson production.  
The theoretical uncertainty of our calculation 
will be estimated through the uncertainty of these initial parameters.
We consider a large enough
longitudinal extension for the deconfined matter: $\eta_0 = \pm 2.2$,
where $\eta_0$ is the space-time rapidity
(this parameter will not influence the transverse momentum spectra at $y=0$).
The constituent quark masses are chosen to be $m_Q=310$ MeV,
$m_S=430$ MeV, $m_C=1540$ MeV.

We calculate the transverse momentum 
spectra of $D$ and $J/\psi$ mesons in Pb+Pb collision
at $T_{had}=175$ MeV and $v_T^0=0.46$.
The $D$ meson is produced from the
coalescence of a charm quark and a ${\overline Q}$ antiquark, while the 
$J/\psi$ is produced as a $C{\overline C}$ state.  
Figure 1 displays the obtained spectra with arbitrary normalization.
The full line shows the MICOR result and the
dashed line indicates the best fit with the parametrization 
\begin{equation}
dN/m_Tdm_T = {C} \cdot m_T^{1/2} \exp(-m_T/T_{eff}) \ 
\label{alpha12}
\end{equation}
in the momentum regions $0.3 \ GeV < m_T - m_i < 2.2 \ GeV$. 
In this region we obtain
$T_{eff}^{Pb+Pb}(D)=259$ MeV and $T_{eff}^{Pb+Pb}(J/\psi)=315$ MeV for
the effective slope parameters. 
Figure 1 shows how the 
non-thermal momentum distribution produced in
the coalescence process (full curve)
mimics the thermalized distribution of eq.~(\ref{alpha12}) 
(dashed curve) in large momentum region. 

The NA38 Collaboration has measured the transverse spectra of the $J/\psi$
in S+U collision at 200 GeV/nucleon bombarding energy \cite{NA38SU}.
To apply our model, we assume that the massive quark-antiquark matter 
was already produced in S+U.
In Figure 1 the experimental data from Ref.~\cite{NA38SU}
are marked by full dots (at arbitrary normalization).
The full curve is the MICOR calculation
using $T_{had}=175$ MeV and a reduced transverse flow,
$v_T^0 = 0.33$. This reduced value for the transverse flow can be
understood, since the smaller project radius of S+U collision 
yields a smaller transverse size and shorter
expansion time for the deconfined phase as compared to 
that in the Pb+Pb collision.
Using the parametrization of eq.~(\ref{alpha12})
in the momentum region $0.3 \ GeV < m_T - m_{J/\psi} < 2.2 \ GeV$
we obtained $T_{eff}^{S+U}(J/\psi)= 234 \ $ MeV (dashed line) 
thus recovered the experimental value,
${\widehat T}_{eff}^{S+U}(J/\psi)= 234 \pm 3$ MeV, obtained in the highest
transverse energy bin (see Ref.~\cite{NA38SU}).
(The published slope value was obtained with the fit 
$dN/m_Tdm_T = C \cdot m_T \cdot K_1(m_T/T)$, 
which is approximately the same one as eq.(\ref{alpha12})
in the temperature region $T \approx 200-300$ MeV 
for particles heavier than the proton,
see Ref.~\cite{MICOR00}.) 
For the D-meson we obtain $T_{eff}^{S+U}(D)= 211 \ $ MeV.
Figure 1 shows the MICOR result (full line),
while the fit from eq.(\ref{alpha12}) is indicated by dashed line.

We now investigate the dependence of the final transverse slope 
on the hadronization temperature and transverse velocity. 
First we fix the hadronization temperature at $T_{had}=175$ MeV 
and vary the transverse flow parameter, $v_T^0$.
Figure 2 shows our predictions
on the effective slopes for the $J/\psi$ (upper curve)
and for the $D$ meson (lower curve) spectra. The open squares show  the
MICOR results in the Pb+Pb collision at 158 AGeV energy
at $v_T^0=0.46$. The uncertainty in the transverse flow, 
$\Delta v_T^0 = \pm 0.05$, 
yields an errorbar on the theoretical predictions:
$T_{eff}^{Pb+Pb}(D)= 259 \pm 19 \ $ MeV, 
$T_{eff}^{Pb+Pb}(J/\psi)= 315 \pm 35 \ $ MeV.
The NA38 result on $J/\psi$ in S+U collision is indicated by
the star and open circles show the MICOR predictions at
$v_T^0=0.33$ . Assuming a 10 \% uncertainty in the transverse flow,
namely $\Delta v_T^0 = \pm 0.03$,  we obtain the theoretical
predictions:
$T_{eff}^{S+U}(D)= 211 \pm 9 \ $ MeV,
$T_{eff}^{S+U}(J/\psi)= 234 \pm 13 \ $ MeV.
These uncertainties show the strong dependence of the charm hadron
slopes on the transverse flow of the intermediate deconfined phase.
As new experimental data will appear, analysis of
Figure 2 can be used to determine the appropriate
$v_T^0$ parameter for the MICOR model  at the
$T_{had}=175$ MeV hadronization temperature.

We repeat this calculation also varying the hadronization 
temperature, $T_{had}$,
with a $\Delta T_{had} = 15$ MeV. The obtained  theoretical
uncertainty remains in the same scale, $\pm 15$ MeV.

In Figure 3, we display the MICOR results for the $T_{eff}$
slopes of the $D$ and $J/\psi$ mesons produced 
in Pb+Pb collisions (they are marked with open squares).
We also display the measured and the calculated slope 
parameters for the $\rho$, $\phi$ and $\Omega$ \cite{MICOR00}.  
Experimental data on the slope parameter of other particles 
are from Refs.~\cite{SQM98WA97,QM99NA50,NA44prl,NA44prot}.
(All data was  parametrized by eq.~(\ref{alpha12})
in the momentum region $m_T-m_i > 0.3$ GeV. ) 

A similar plot can be created for the S+A collisions at 200 GeV/nucleon
project energy. Figure 4. includes
results for S+U \cite{NA38SU}, S+W \cite{WA85a,WA85d} and
S+Au \cite{NA35a} collisions. The open squares show the
MICOR results for particles $\rho$, $\phi$, $\Omega$, $D$ and $J/\psi$.
Here the $J/\psi$ data determines the transverse flow parameter, 
$v_T^0=0.33$, which reproduce the slope for the $\phi$ meson.

Figures 3 and 4 support the assumption that the secondary
hadronic interactions only slightly modify the transverse momentum spectra
at higher momenta, $m_T - m_i > 0.3$.
They  reveal that the separation of the $p^+$, ${\overline p}^-$, $\Lambda$,
${\overline \Lambda}$, $\Xi^-$ and ${\overline \Xi}^+$
from the weakly interacting $\phi$ and $\Omega$ is small.
Even more, the slopes of the $p^+$ and ${\overline p}^-$
are very close to that of the weakly interacting $\phi$ meson. 
Thus we do not expect considerable modification
of the much heavier charm mesons transverse spectra.

Our result sides with the  analysis of hydrodynamical
expansion versus hadronic cascade evolution performed by URQMD \cite{URQMD},
where enhanced initial transverse flow was paired with a reduced influence
of hadronic cascading in the transverse momentum spectra of heavier particle.
In this manner, our result does not contradict to recent finding of
Lin et al. \cite{LinKoZ}, where relatively large modification was found
from "hadron+D" collisions of the D-meson spectra in a longitudinally 
expanding system. Allowing a relatively large transverse flow, e.g.
$v_T^0=0.46$ as in MICOR, the lifetime of the fireball will be shortened
and the efficiency of the hadronic rescattering will be reduced.
\bigskip

In this paper, we predict the transverse slope of the $D$ and
$J/\psi$ mesons in Pb+Pb collision at CERN SPS  energy.
Using the microscopical MICOR model we could follow the hadronization
of the quark matter that we assume is produced in this
heavy ion collision, and calculate the transverse momentum distributions.
We obtained 
$T_{eff}^{Pb+Pb}(D)= 259 \pm 19 \ $ MeV and
$T_{eff}^{Pb+Pb}(J/\psi)= 315 \pm 35 \ $ MeV.
We compare our model with experimental data on $J/\psi$ production
in S+U collision and demonstrate that we can reproduce the measured spectra
with a reasonably  reduced transverse flow of the quark matter
at a hadronization temperature of $T_{had}=175$ MeV.
We predicted  the slope parameter of the D-meson to be
$T_{eff}^{S+U}(D)= 211 \pm 9 \ $ MeV.
We urge future experiments to measure this quantity to provide further 
tests of this model.

\bigskip

{\bf Acknowledgment:}
We thank M. Gyulassy and 
J. Zim\'anyi for stimulating discussions. 
One of the authors (P.L.) is especially grateful to S.E. Vance for 
his comments and his careful reading of the manuscript.
This work was supported by the OTKA Grant No. T025579,
the US-Hungarian Joint Fund No. 652, 
and partly by the DOE Research Grant under Contract No.
de-FG-02-93ER-40764.

\newpage
\begin{center}
\vspace*{18.0cm}
\includegraphics{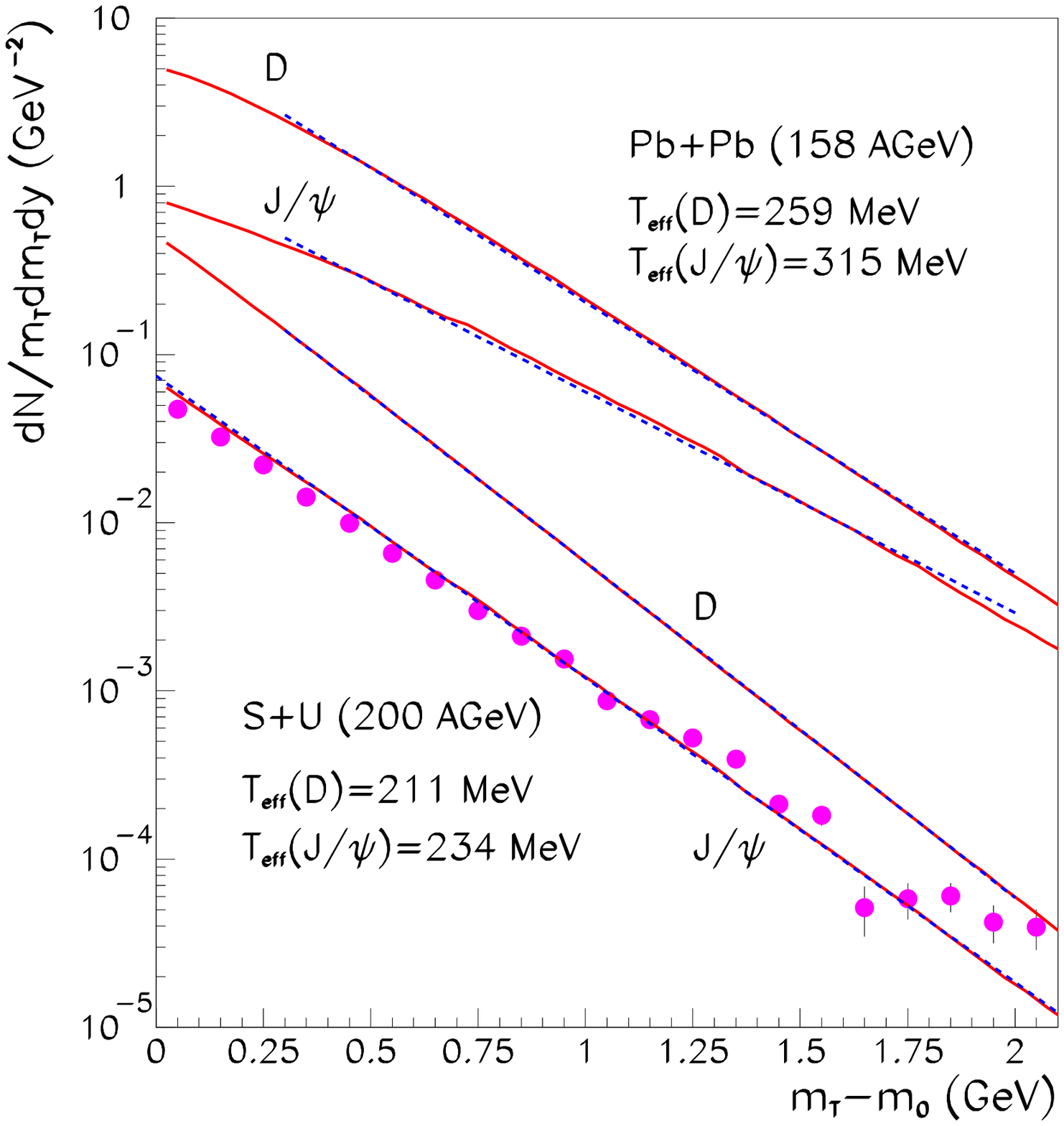}
\begin{minipage}[t]{15.cm}
{ {\bf Fig.~1.}
The transverse momentum spectra of the $D$ and $J/\psi$ meson
calculated by the MICOR model with parameters 
$T_{had}=175$ MeV and $v_T^0=0.46$ for Pb+Pb collision at 158 AGeV 
and with $v_T^0=0.33$ for S+U collision at 200 AGeV (solid lines). 
The dashed lines are the fits of eq.(\ref{alpha12})
in the transverse momentum region
$0.3 < m_T - m_i < 2.2$ GeV.
The full dots are the data from the NA38
Collaboration for S+U collision \cite{NA38SU}.
}
\end{minipage}
\end{center}

\newpage
\begin{center}
\vspace*{18.0cm}
\includegraphics{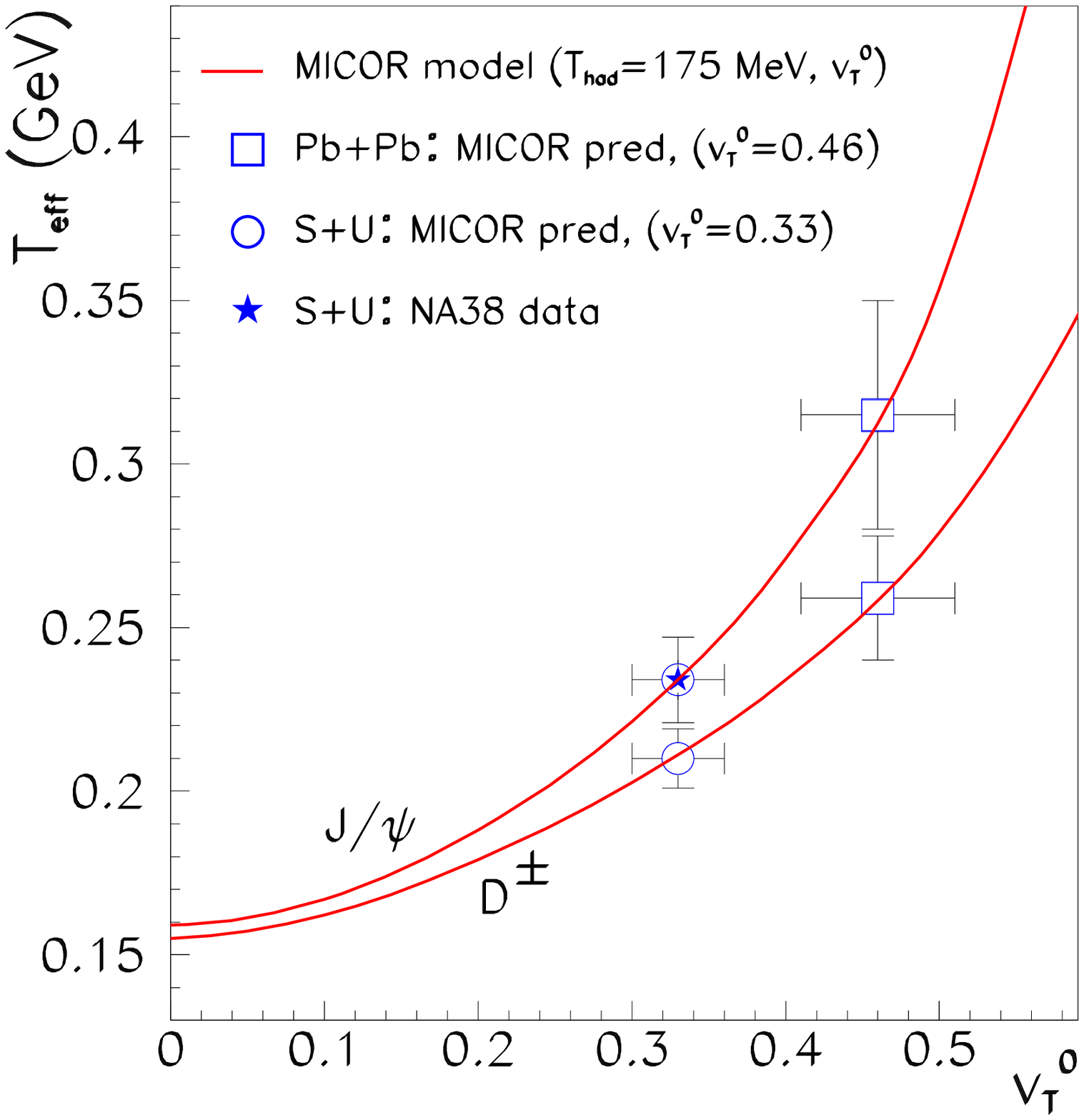}
\begin{minipage}[t]{15.cm}
{ {\bf Fig.~2.}
Prediction from the MICOR model for
the $T_{eff}$ slopes of the $D$ and $J/\psi$ transverse momentum spectra
as the function of initial flow $v_T^0$ at $T_{had}=175$ MeV.
The open squares indicate the MICOR prediction with $v_T^0=0.46$
for Pb+Pb collisions at 158 AGeV and the open circles indicate the
prediction with $v_T^0=0.33$ for the S+U collisions at 200 AGeV.
The errorbars shows the influence of 10 \% uncertainty in the
transverse flow.
The star represents the NA38 experimental data 
for the $J/\psi$ slope
in S+U collisions,
see Figure 1 and Ref.~\cite{NA38SU}.
}
\end{minipage}
\end{center}

\newpage
\begin{center}
\vspace*{18.0cm}
\includegraphics{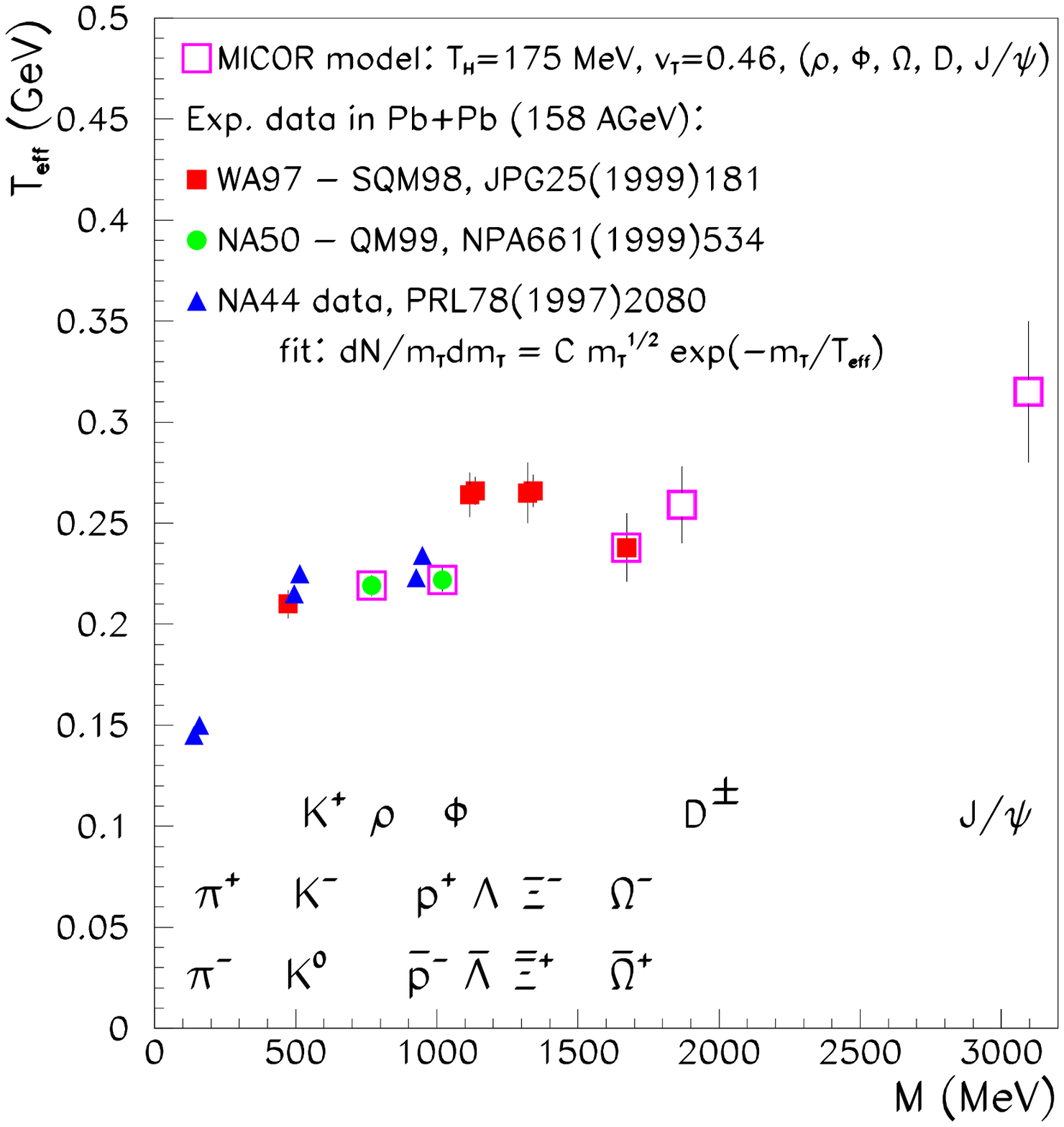}
\begin{minipage}[t]{15.cm}
{ {\bf Fig.~3.}
Experimental hadronic slopes of the transverse momentum spectra 
in the PbPb collision at 158 AGeV energy from WA97 \cite{SQM98WA97} (squares),
NA50 \cite{QM99NA50} (dots) and NA44 Collaboration \cite{NA44prl}
(triangulars).
The results of WA97 and NA50 were originally fit by eq.(\ref{alpha12}).
We fit the NA44 data on $\pi^\pm$,
$K^\pm$, $p^+$ and ${\overline p}^-$ \cite{NA44prl} 
in the same way in the momentum region
$m_T-m_i > 0.3$ GeV.
Open squares indicate the MICOR results on charm mesons and other
hadrons.
}
\end{minipage}
\end{center}

\newpage
\begin{center}
\vspace*{18.0cm}
\includegraphics{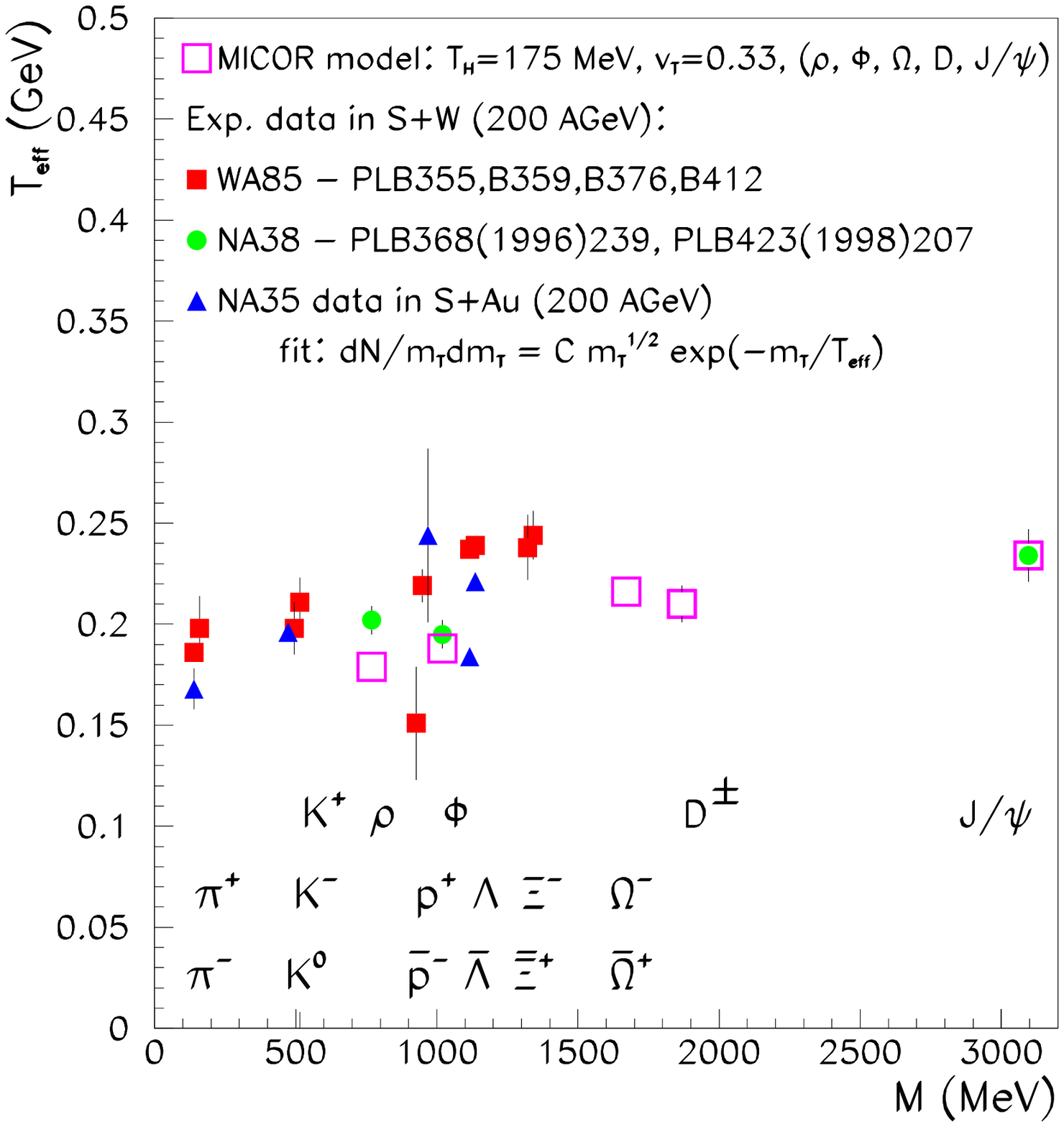}
\begin{minipage}[t]{15.cm}
{ {\bf Fig.~4.}
Experimental hadronic slopes of the transverse momentum spectra 
in the S+W and S+Au collisions at 200 AGeV energy from WA85 
\cite{WA85a,WA85d} (squares),
NA38 \cite{NA38SU,NA38SUrho} (dots) and NA35 Collaboration \cite{NA35a}
(triangulars).
The results of WA85 and NA38 data were originally fit by eq.(\ref{alpha12}).
We fit the NA35 data on $\pi^-$,
$K^0_S$, $p^+$, $\Lambda$ and ${\overline \Lambda}$ \cite{NA35a} 
in the same way in the momentum region
$m_T-m_i > 0.3$ GeV.
Open squares indicate the MICOR results on charm mesons and other
hadrons.
}
\end{minipage}
\end{center}

\end{document}